\documentclass[12pt]{article}
\usepackage[dvips]{epsfig}
\begin{document}

\title{Polyelectrolyte-induced peeling of charged multilamellar vesicles}
\author{Edith Vivares and Laurence Ramos\thanks{E-mail: ramos@gdpc.univ-montp2.fr}
\\
Laboratoire des Collo\"{i}des, Verres et Nanomat\'{e}riaux (UMR
CNRS-UM2 5587)\\ CC26, Universit\'{e} Montpellier 2, 34095
Montpellier Cedex 5\\ France}


\date{\today}

\maketitle

\begin{abstract}

We study mixtures of charged surfactants, which alone in solution
form uni- and multilamellar vesicles, and oppositely charged
polyelectrolytes (PEs). The phase behavior is investigated at
fixed surfactant concentration as a function of the
PE-to-surfactant charge ratio $x$. We find that, for $x>0$,
aggregates form. Light microscopy and X-ray scattering experiments
show that the isoelectric point plays a crucial role since the
morphology and the microscopic structure of the aggregates are
different before ($x\leq1$) and after the isoelectric point
($x>1$). To better understand the dynamics for the formation of
PE/surfactant complexes, we perform light microscopy experiments
where we follow in real-time the effect of a PE solution on one
multilamellar vesicle (MLV). We find that the PE induces a peeling
of the bilayers of the MLV one by one. The peeling is accompanied
by strong shape fluctuations of the MLV and leads ultimately to a
pile of small aggregates. This novel phenomenon is analyzed in
detail and discussed in terms of PE-induced tension, and pore
formation and growth in a surfactant bilayer.

\end{abstract}

\pagebreak


\section{Introduction}

The phase behavior of mixtures of polymers and surfactants is
extensively studied because of both fundamental and industrial
interests \cite{PolymSurf}, these mixed systems being widely used
in industrial applications ranging from food, to cosmetic or
health-care. Among polymer/surfactant composites, blends of
polyelectrolyte (PE) and oppositely charged surfactants have
received particular attention \cite{PESurf}. Very generally,
around the isoelectric point, which is characterized by an equal
number of positive and negative charges brought by the two
species, an associative precipitation occurs. The driving force
for this associative precipitation is mainly the electrostatic
interactions between PE and surfactants where entropy is gained
thanks to the counterions released upon polymer-surfactant
binding. Other forces, such as the hydrophobic interactions
between the polyelectrolyte backbone and the surfactant
hydrocarbon tails \cite{hydrophobic}, may however play an
important role as well. Recently, it has been shown that the
condensate exhibits very often a long-range order reminiscent of
the structures classically found in surfactant/water mixtures
(lamellar, hexagonal and cubic phases)
\cite{PE_TA1,PE_TA2,PE_TA3}. When the surfactant forms bilayers,
the presence of PE may convert the largely separated bilayers into
densely packed membrane stacks with PE strands sandwiched between
the surfactant bilayers. This phenomenon has been referred to as
bridging by the polyelectrolyte \cite{bridging1, bridging2} and
appears promising for many industrial or therapeutic applications.
However, its exact mechanism and its dynamics have never been
investigated experimentally.

Thus, despite a large amount of experimental studies, surprisingly
enough, not much is known on the dynamics of associative
precipitation between PE and oppositely charged surfactants. In
order to gain insight on the dynamics, novel experimental
approaches are needed. We therefore have started real-time
observations by light microscopy of the effect of a
polyelectrolyte solution on assemblies  of oppositely charged
surfactants. Our experiments allow a detailed investigation of the
polyelectrolyte-induced morphological transition of a
multilamellar vesicle (MLV) to be performed. They show quite
unexpectedly that the polyelectrolyte provokes a slow but
continuous peeling of the MLV through successive removal of the
bilayers one after the other.

The paper is organized as follows. In Sec.~\ref{SEC:phase_diag},
we investigate by visual inspection, light microcopy, and X-ray
scattering the phase-diagram of the PE/surfactant mixtures
obtained by gently mixing together the PE and the surfactant
solutions. In the range of surfactant concentration investigated,
we find that the structure and morphology of the PE/surfactant
complexes depend only on the PE-to-surfactant charge ratio, $x$.
We then present in Sec.~\ref{SEC:peeling} our main results
concerning real-time observation of the interaction between a MLV
and a PE solution and finally discuss a possible physical
mechanism for polyelectrolyte-induced peeling of the MLVs. We
conclude in Sec.~\ref{SEC:conl}.


\section{Phase diagram of the surfactant/polyelectrolyte mixtures}
\label{SEC:phase_diag}

\subsection{Materials}

We use Didodecylammonium Bromide, DDAB,
$(C_{12}H_{25})_{2}N^+(CH3)_2Br^-$, as cationic surfactant and an
alternating copolymer of styrene and maleic acid in its sodium
salt form, $-CH_2CH(C_6H_5)CH(CO_2Na)-CH(CO_2Na)-$, as negatively
charged polyelectrolyte. The polyelectrolyte (PE) and DDAB are
purchased from Aldrich and used as received. The PE molar weight
is $120 000$ g/mol, which corresponds approximatively to $500$
monomers per molecule. The pH of the PE solution is around $9$,
while the pKa of the two acidic functions of maleic acid  are
$1.8$ and $6.1$; this ensures that the PE is indeed in the basic
form as indicated by the supplier and possesses thus $2$ negative
charges per monomer.

We call $C_S$ and $C_{PE}$ respectively the surfactant and
polyelectrolyte concentrations. In most experiments, we use dilute
surfactant solutions with $C_S=3$ mM or $C_S=22$ mM (it
corresponds respectively to $0.14$ and $1$  \% w/w). At these
concentrations, surfactant molecules simply diluted in water give
a mixture of unilamellar vesicles and multilamellar vesicles
(MLVs), with a relatively large amount of big MLVs (size $\geq10\,
\mu \rm{m}$). We define $x$ as the polyelectrolyte-to-surfactant
charge ratio ($x=1$ corresponds to the stoichiometric charge
neutrality) and calculate $x$ assuming that there is no counterion
condensation. As we will see later, different behaviors are
observed in the phase diagram, for $x \leq 1$ and $x>1$. The fact
that the cross-over between two regimes is observed at $x=1$ is a
clear hint that our determination of $x$ should be correct. In
most experiments, $x$ varies between $0$ and $50$, which
corresponds to a maximum PE concentration $C_{PE}$ equal to $2.0
\%$ w/w (resp. $14.5 \%$ w/w) for $C_S=3$ mM (resp. $C_S=22$ mM)

\subsection{Light microscopy observations}

Samples are prepared by gently mixing together a surfactant
solution  and a polyelectrolyte solution. The final surfactant
concentration is fixed ($C_S=3$ mM or $C_S=22$ mM) and the PE
concentration is changed in order to vary $x$.

Macroscopically, one finds that as PE is added to the surfactant
solution, the mixture becomes instantaneously turbid, indicating
the presence of aggregates of micrometer size. The turbidity
continuously increases as the amount of PE increases up to the
isoelectric point. Above the isoelectric point ($x>1$), aggregates
precipitates into larger flocculates that may stick to the
container walls. As more PE is incorporated, the flocculates start
to resolubilize until a totally transparent solution is recovered
at very large PE excess (for $x$ larger than $\approx 50$).

\begin{figure}
\includegraphics{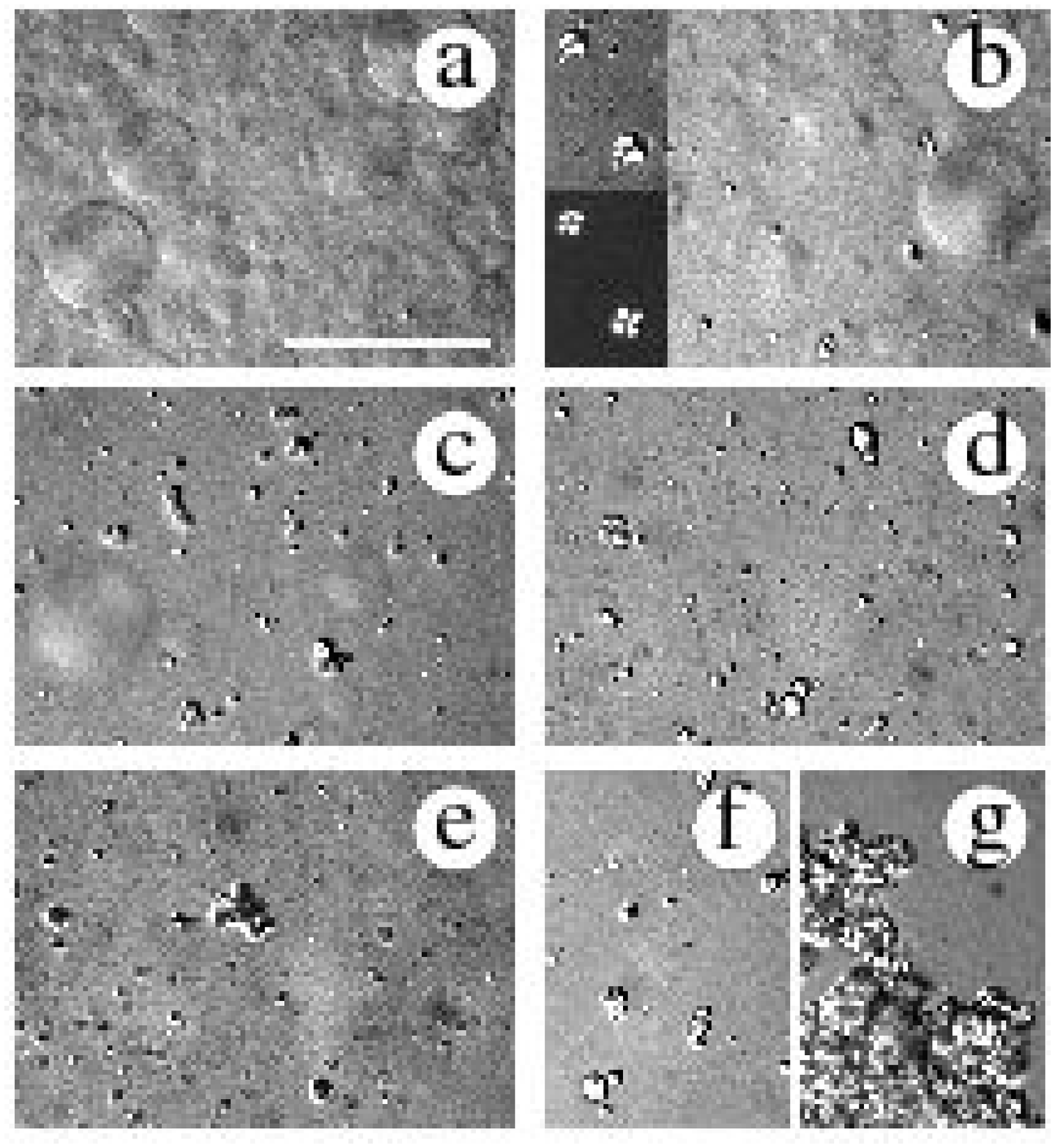}
\caption{Light microscopy pictures of mixtures of DDAB and PE. The
surfactant concentration is fixed at $3$ mM. The PE-to-surfactant
charge ratio is $x=$ (a) $0.02$, (b) $0.2$, (c) $0.5$, (d) $1$,
(e) $2$, (f,g) $10$. All pictures are observed using differential
interference contrast except the left hand side (bottom) of (b)
where the sample is observed between crossed polarizers. The field
of view of each of the two images at the left hand side of (b) is
$29 \, \mu\rm{m} \times 44 \, \mu\rm{m}$ ; the scale is the same
for all other images ; the white bar corresponds to $50\,
\mu\rm{m}$.} \label{FIG:1}
\end{figure}

Observations of the samples with a light microscope corroborates
the macroscopic observations described above. Figure \ref{FIG:1}
presents a series of micrographs, taken under differential
interference contrast conditions, of the PE/surfactant systems
that follow the evolution of the microstructures with changes in
the PE-to-surfactant charge ratio. For $x<1$, that is in excess of
surfactant, one observes the coexistence of apparently unaffected
surfactant assemblies and aggregates (fig.\ref{FIG:1}a-c). Even
for $x$ very small, hence in large excess of surfactant, one
observes aggregates instead of a solubilization of the PE by the
surfactant as very often seen in PE/surfactant mixtures (a similar
effect has been noticed in Ref.\cite{Talmon}). We note that the
aggregates can be unambiguously distinguished from the
multilamellar vesicles because of their much higher contrast. The
aggregates have a rather spherical shape and a smooth interface.
They are polydisperse with a maximum size of about $5 \, \mu$m.
Moreover, they  are birefringent and exhibit surprisingly nice
maltese crosses when viewed between crossed polarizers (see
fig.\ref{FIG:1}b), suggesting that their structure is lamellar
\cite{Maltese}. The respective proportion of MLVs and aggregates
decreases as $x$ increases and at the isoelectric point, no
surfactant multilamellar vesicles are apparently present anymore:
only the spherical and birefringent aggregates persist
(fig.\ref{FIG:1}d). Beyond the isoelectric point, in excess of PE,
large flocculates with more corrugated surface than the $x<1$
aggregates are observed (fig.\ref{FIG:1}e-g). These flocculates
eventually dissolve in a large excess of polyelectrolyte (for
$x\simeq50$). They are also birefringent but contrary to the
aggregates observed below the isoelectric point, no clear texture
is distinguishable between crossed polarizers.

We note that all the structures described here form
instantaneously upon mixing the PE and surfactant solutions and
appear to be quite stable, as they do not evolve with time.

\subsection{X-ray scattering experiments}

Small-angle X-ray scattering (SAXS) experiments have been
performed  using an in-house setup with a rotating anode X-ray
generator equipped with two parabolic mirrors giving a highly
parallel beam of monochromatic Cu K$\alpha$ radiation (wavelength
$\lambda=0.154$ nm). The SAXS intensity is collected with a
two-dimensional detector.

\begin{figure}
\includegraphics{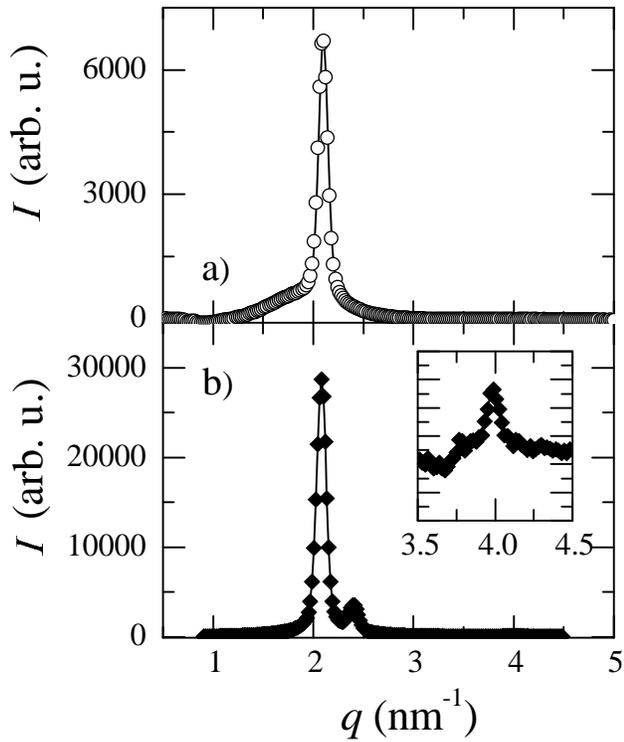}
\caption{Small-angle X-ray scattering patterns of samples with (a)
$C_S=108$ mM and $x=1$, (b) $C_S=22$ mM and $x=3.5$. Inset is a
blow-up of the large $q$ region.} \label{FIG:2}
\end{figure}

We find that the SAXS patterns of the mixtures of PE and
surfactant do not depend on the surfactant concentration $C_s$
(for $C_s$ in the range $3-110$ mM). Two markedly different
scattering patterns are however obtained for $x \leq 1$, that is
in excess of surfactant or at the isoelectric point, and for
$x>1$, that is in excess of PE. A typical spectrum measured in the
$x\leq 1$ regime is shown in fig.\ref{FIG:2}a. A unique and sharp
peak is obtained at a wave-vector $q_0=2.1 \, \rm{nm}^{-1}$. This
peak is unambiguously due to the scattering of the aggregates
aggregates because at equivalent concentrations the scattered
intensity of the surfactant or of the polyelectrolyte alone in
solution is essentially flat in the range of q-vectors
investigated here ($0.5-5 \, \rm{nm}^{-1}$). Although the
birefringence pictures of the aggregates suggest that they have a
lamellar symmetry, no peak can be detected at higher wave-vector
(in particular at $2q_0$). The spectra shown fig.\ref{FIG:2}a can
be compared to the one obtained when a high salt (NaCl)
concentration ($1$M) is added to a same DDAB solution causing the
collapse of the surfactant bilayers: in that case, two peaks are
obtained, a sharp and intense one at a wave-vector slightly
smaller than the one obtained for the DDAB/PE mixture ($q*=2.0 \,
\rm{nm}^{-1}$) and a weak second order at $2q*$ (with maximum
intensity more than one order of magnitude smaller than the one of
the first peak), indicating a condensed lamellar phase. Because of
the birefringence signature, as well as the fact that for the
PE/DDAB mixture the full width at half maximum of the peak is
equal to the one of the first order of the lamellar phase induced
by NaCl, we believe that the PE-induced aggregates possess also a
lamellar symmetry. However, because of contrast effect, the second
order could not be detected because it would fall in the close
vicinity of a minimum of the form factor. To conclude, the
PE-induced aggregates are presumably PE/surfactant complexes
constituted of a condensed lamellar phase with an interlayer
spacing  $d=2 \pi /q_0 \simeq 3.0$ nm, which is slightly larger
than the surfactant bilayer thickness $\delta=2.3$ nm
\cite{DDAB1,DDAB2}.

Beyond the isoelectric point, the microscopic structure is
radically different. Three peaks are systematically obtained at
$q_0= 2.1 \, \rm{nm^{-1}}$, $q_1= 2.4 \, \rm{nm^{-1}}$ and $q_2=
4.0 \, \rm{nm^{-1}}$. As can be seen in fig.\ref{FIG:2}b, the
second peak has a maximum intensity about one order of magnitude
smaller than the first one, while the third one is extremely weak,
with a maximum intensity more than two orders of magnitude smaller
than that of the first peak. The relative positions of the three
peaks rule out a hexagonal symmetry ;  a cubic symmetry is also
excluded because aggregates are birefringent. In addition, because
the relative intensities of the peaks at $q_1$ and $q_2$ does not
change when varying $x$ (but keeping $x>1$), these two peaks do
not originate from the coexistence of two types of aggregates in
our samples but correspond instead to a unique structure. Thus,
the data suggest rather a complex structure with a condensed
lamellar phase, giving rise to the two peaks at $q_0$ and
$q_2\simeq 2q_0$, with an in-plane order, which is at the origin
of the peak at $q_1$. The interlamellar distance is $d=2\pi/q_0
\simeq 3.0$ nm and the characteristic size of the in-plane order
is $2\pi/q_1 \simeq 2.6$ nm. We note that the interlayer spacing
is the same as the one observed before the isoelectric point, the
only difference between the structure before and after the
isoelectric point being the existence of the in-plane order.

\subsection{Discussion}

Both light microscopy and X-ray scattering evidence two distinct
regions, depending on the PE-to-surfactant charge ratio. One
important result is that only two structures exist for the
PE/surfactant complexes. In particular the structural
characteristic lengths do not depend on the surfactant
concentration neither on the values of the global PE-to-surfactant
charge ratio of the mixtures $x$ but only on whether $x$ be
smaller or larger than $1$. This suggests that each of the two
types of PE/surfactant complexes is characterized by a fixed
PE-to-surfactant charge ratio : before the isoelectric point,
because of the depletion in PE, complexes and apparently
unaffected surfactant self-assemblies coexist in solution, and
after the isoelectric point, the excess of PE is used to
resolubilize the aggregates, the mixture being composed of
resolubilized species (which are not observable by light
microscopy and do not give a signal in the wave-vectors window
probed by SAXS) and PE/surfactant complexes, directly observed by
light microscopy.

Our results suggest moreover the existence of an in-plane
correlation. We note that this type of structure is reminiscent of
the one measured when DNA is added to a mixture of neutral and
cationic lipids. In that case, the lamellar condensates of lipids
and DNA shows a two-dimensional in-plane order of DNA intercalated
between the lipidic bilayers \cite{DNA1,DNA2}. Contrary to what
has been measured with DNA, the in-plane correlation peak measured
here appears only in excess of polyelectrolyte and its position is
independent of the PE-to-surfactant charge ratio $x$ (as long as
$x>1$). A crucial difference between the experimental conditions
of Refs.\cite{DNA1,DNA2} and ours is presumably that DNA is a much
stiffer polyelectrolyte than the one we use. In-plane correlation
has also been observed by electron microscopy for flexible
polyelectrolytes \cite{inplanecorrelation} and has been attributed
to hexagonally ordered polyions blobs at the surface of a lipid
bilayer.  No quantitative investigation on the concentration
dependence of the in-plane correlation length has however been
performed in that case.

In summary, the structures of the PE/surfactant complexes both
before and after the isoelectric point are in agreement with the
physical picture of bridging of the bilayers by the PE molecules
since a condensed lamellar phase is measured in both cases. In
addition, because the PE we use is an alternated copolymer of a
charged monomer and a hydrophobic monomer (styrene), insertion of
the PE in the surfactant  bilayers is conceivable. The rather
peculiar features of the phase diagram described here may be due
to a subtle interplay between electrostatic and hydrophobic
interactions between the surfactant bilayers and the
polyelectrolyte molecules.


\section{Real-time observations: polyelectrolyte-induced peeling of multilamellar vesicles}
\label{SEC:peeling}

In order to gain insight on the dynamics of the condensed lamellar
phase, we perform time-resolved light microscopy experiments where
we follow in real-time the morphological transitions of a MLV
induced by a polyelectrolyte solution.

\subsection{Results}

\subsubsection{Experimental procedure}

\begin{figure}
\includegraphics{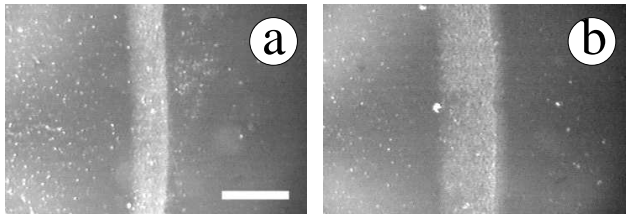}
\caption{(a,b) Light microscopy picture of the interface between a
PE solution and a DDAB solution. The pure PE (resp. DDAB) solution
stands at the left (resp. right) hand-side of the pictures. The
time elapsed between the two images is $330$ s. The scale is the
same for the two images; the white bar corresponds to $0.5$ mm.}
\label{FIG:3}
\end{figure}

We use an observation chamber of approximative thickness $100 \,
\mu$m formed from heat sealing a glass coverslip to a glass slide
with a Parafilm spacer. The surfactant solution is introduced by
capillarity into the chamber, a drop of the PE solution is then
put into contact with the surfactant solution, after which the
apertures of the chamber are sealed with vacuum grease to avoid
solvent evaporation and convective flow. In this way, a nice and
smooth interface between the surfactant and the PE solutions is
formed. The interface is betrayed by a strong contrast in bright
field light microscopy because of the presence of highly
contrasted (with respect to pure PE solution and to pure
surfactant solution)  PE/surfactant complexes (fig.\ref{FIG:3}).
As interdiffusion of the species proceeds, the interface initially
very thin broadens.

We fix the surfactant and PE concentrations respectively at $0.14$
\% w/w and $0.46$, which correspond to concentrations of electric
charges of respectively $22$ mM and $35$ mM. The concentration and
respective volume of the two solutions are such that the overall
PE-to-surfactant ratio is of the order of $1$. Note that we have
diluted the PE in a mixture of water and glycerol ($29 \% $ water)
so as to increase the viscosity of the PE solution and hence slow
down the diffusion of the PE into the surfactant solution
\cite{Note_glycerol}.

In most experiments we focus on a unique large MLV using
differential interference microscopy. Because of its large size,
the MLV is essentially immobile in the field of view. The
PE-induced morphological transitions are recorded on video-tapes
and digitalized afterwards. We point out that in this type of
experiment out-of-equilibrium structures are observed since there
is a continuous supply of polyelectrolyte due to diffusion.

\subsubsection{Observations}

\begin{figure}
\includegraphics{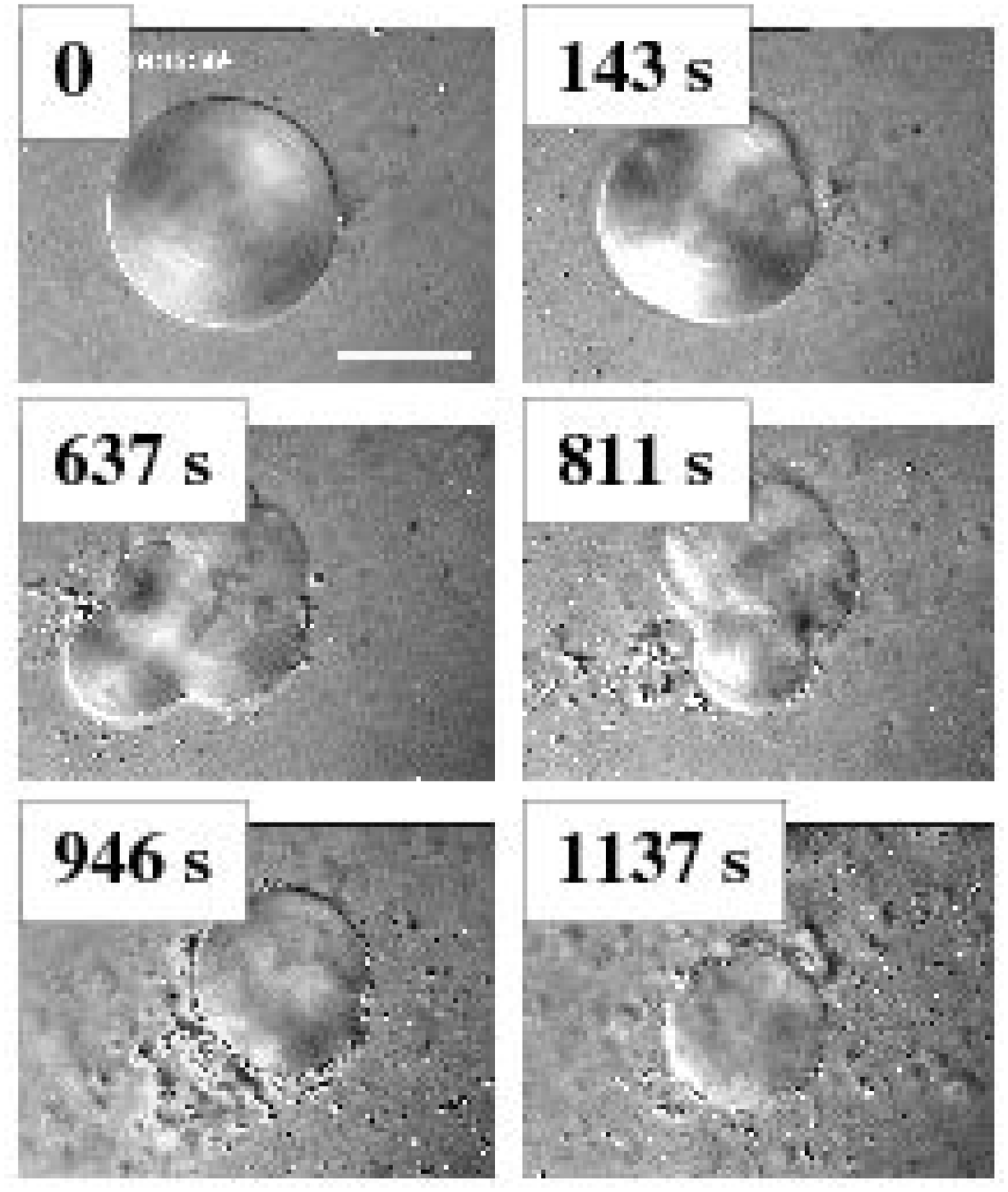}
\caption{Differential interference contrast optical micrographs of
morphological changes a MLV due to its interaction with a PE
solution. The time lapse is indicated on each image. The scale is
the same for all images; the white bar corresponds to $50\,
\mu\rm{m}$.}
 \label{FIG:4}
\end{figure}

\begin{figure}
\includegraphics{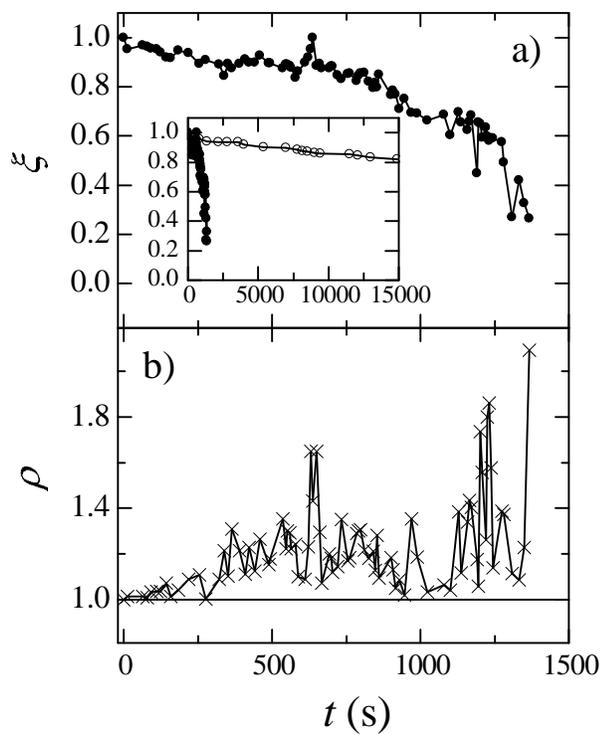}
\caption{Time evolution of the (a) average size normalized by the
initial size, and (b) anisotropy ratio, of the MLV shown in fig.
\ref{FIG:5}. Inset: Black circles are the same data as in the main
plot and open circles correspond to the size decrease of a MLV in
contact with a neutral polymer (POE of molar weight $1000000$
g/mol at $17.5$ w/w \% in water).} \label{FIG:5}
\end{figure}

Figure \ref{FIG:4} presents a series of snapshots of a MLV as it
interacts with the polyelectrolyte solution. Time zero in this
experiment is defined as the time where the MLV starts to
experience modifications. This time coincides with the time at
which small aggregates with a strong contrast appear in the field
of view: this is the time when the PE, which is diffusing, reaches
the MLV. As time goes on, one observes that the size of the MLV
decreases continuously until its complete disappearance. The size
decrease is accompanied by the formation in the close vicinity but
also at the surface of the MLV of small aggregates A pile of small
aggregates is left after complete disappearance of the MLV. The
aggregates have a strong contrast and a morphology analogous to
the PE/surfactant complexes observed in the regime above the
isoelectric point (excess of PE) (fig.\ref{FIG:1}e-g). A
remarkable feature is that the size decrease and the formation of
aggregates are associated with marked shape distortion of the
initially spherical MLV. The shape of the MLV fluctuates strongly
and can evolve very rapidly from a very distorted shape to a
perfectly spherical one and back to another distorted shape. In
order to quantify the morphological transition of the MLV, we
define the average size of the MLV as the arithmetic mean of the
minor, $l$, and major, $L$, dimensions of the MLV as measured from
digitalized micrographs of $2$D cross-sections of the MLV. In
fig.\ref{FIG:5}a, we plot the time evolution of this quantity
normalized by the initial size of the MLV, $\xi$. We observe a
continuous decrease of $\xi$, which has diminished by half in
about $1200$ s. In addition, we quantify the shape distortions by
the anisotropy ration $\rho=L/l$, which is found to range from $1$
(for a spherical object, see for instance fig.\ref{FIG:4} at $t=0,
946$ and $1137$ s) and almost $2$ (when the MLV is highly
distorted as in fig.\ref{FIG:4} at $t=637$ and $811$ s). As can be
seen in fig.\ref{FIG:5}b, $\rho$ varies in a very erratic way with
time, reflecting the fast changes of shape of the MLV.

We believe that the polyelectrolyte induces a progressive peeling
of the MLV, i.e. a discrete removal of the bilayers of the MLV one
after the other. Several experimental observations support this
mechanism. First of all, all along the process and until the
complete disappearance of the MLV, the contrast between the MLV
and the outside solvent does not change and the contrast inside
the MLV remains homogeneous. This suggests that all phenomena take
place only at the interface between the MLV and the solvent as
expected for a peeling mechanism for which, at any time, only the
outer bilayer (i.e. the one in contact with the outside solvent)
interacts with the polyelectrolyte solution.

\begin{figure}
\includegraphics{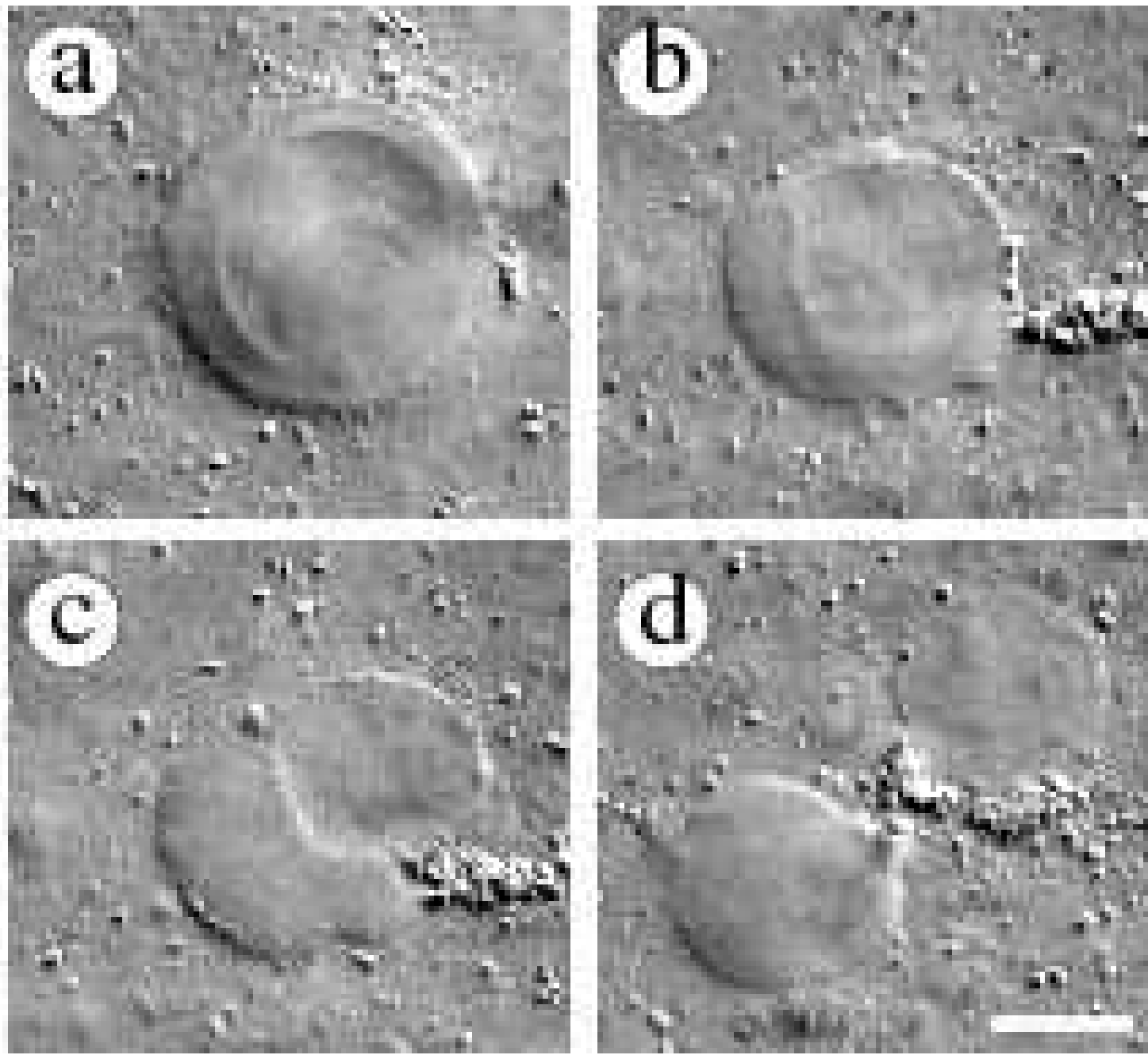}
\caption{Differential interference contrast optical micrographs
showing the scission of a MLV into two provoked by a PE solution.
From left to right $t=0$, $2$ min $1$ s, $2$ min $4$ s and $2$ min
$23$ s. The scale is the same for all images ; the white bar
corresponds to $20\, \mu\rm{m}$.} \label{FIG:6}
\end{figure}

Moreover, we were able to observe occasionally the division of one
vesicle into two, which subsequently evolved independently. A very
naive interpretation of this observation, compatible with the
peeling scenario, is that at a certain time of the evolution of
the mother MLV, the two preexisting daughters vesicles were
uniquely maintained together by one bilayer. Once this bilayer has
been peeled off, the two MLVs are released in solution and become
independent one with respect to the other one. We point out that
the micrographs showing this event (fig.\ref{FIG:6}) are
reminiscent to what has been recently observed in
Ref.\cite{scission} but the physical mechanisms involved are
different since membrane is created during the process described
in Ref.\cite{scission}.

\begin{figure}
\includegraphics{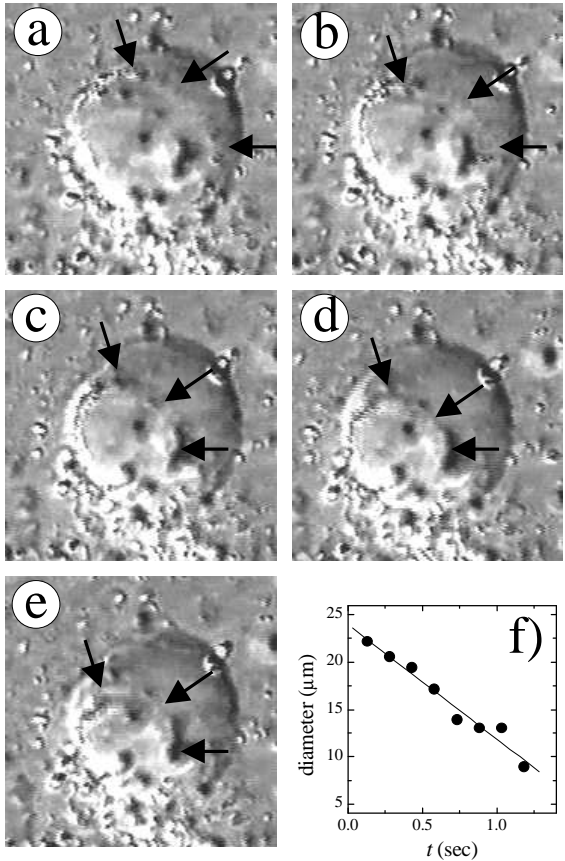}
\caption{(a,e) Micrographs showing one peeling event. Black arrows
help to visualize the retraction of the rim. Images are separated
by $150$ ms; field of view: $40\, \mu\rm{m} \times \, 40\,
\mu\rm{m}$. (f) Plot of the diameter of the rim as a function of
time ($t=0$ correspond to the time where the rim diameter equals
the vesicle diameter.) ; symbols are experimental data and line is
a linear fit yielding a retraction speed $V=12\, \mu$m/s.}
\label{FIG:7}
\end{figure}

Finally, a careful inspection of the video-tapes allows individual
peeling events to be identified. The series of micrographs in
fig.\ref{FIG:7} shows several steps of one peeling event. In these
pictures, a rim decorated by small aggregates is visualized, whose
initial diameter is as large as the diameter of the MLV and which
retracts with time until complete closure. The retraction lasts
less than $1.5$ s and occurs as constant speed $V$ as can be shown
from the plot of the radius of the rim as a function of time
 (fig.\ref{FIG:7}f). We find  $V=12.1 \pm 0.9 \, \mu \rm{m}/s$.
 The closing of the rim marks
the late stage of one peeling event that can be schematically
described as follows: rupture of the outside bilayer, opening-up
and growth of a hole until failure (i.e. detachment of the
bilayer). Note that beyond the point when the diameter of the hole
is larger than that of the MLV, the growth of the hole is
visualized as the retraction (in a point diametrically opposed to
the opening-up location) of the remaining bilayer. The retraction
of the bilayer is visualized in fig.\ref{FIG:7}. We note that we
were not able to experimentally track the
 opening up of a hole
which implies that this first stage must be much more rapid than
the second stage observed experimentally.

A simple estimation of the size of the PE/surfactant complexes can
be done, assuming that each peeled-off bilayer gives birth to one
aggregate. The peeling of one bilayer of initial diameter $D$
releases a surface area $\pi D^2$, which will entirely be used to
create a condensed aggregate. For the sake of simplicity, the
aggregate is assumed to be a cube of side $r$ constituted of a
perfectly lamellar structure with a periodicity $d=3\, \rm{nm}$
equal to that measured by SAXS. Equalling the surface areas reads
$\pi D^2 = \frac{r^3}{d}$, hence $r = (\pi D^2 d)^{1/3}$. For the
MLV shown in fig.\ref{FIG:4}, $D=90\,\mu \rm{m}$ at the beginning
of the process ($t=0$). One thus obtains $r \simeq 4 \,
\mu\rm{m}$, in good agreement with the size of the complexes
measured in the vicinity of the MLV a few minutes later. This
evaluation implies that no multiple pores are formed
simultaneously on one given bilayer, which would result in smaller
complexes. In addition, one would also expect the size of the
complexes to decrease as the peeling proceeds since smaller
surface of bilayer is available as the peeling progresses into the
core of the MLV. However, testing this conjecture is difficult, in
particular because, as mentioned above, all PE/surfactant
complexes formed do not originate from the bilayers peeled-off
from large MLVs but also from small surfactant assemblies,
undetectable by light microscopy.

\subsection{Discussion}

\subsubsection{Role of osmotic effects?}

As a control experiment, we use the same experimental procedure
but take a neutral polymer (polyoxyethylene with a molar weight of
$100000$ g/mol, at a concentration of $17.5\%$ in water) in place
of the polyelectrolyte. When a MLV is put in contact with the
neutral polymer solution, its size decreases with time, due to an
osmotic de-swelling, but the MLV keeps a perfectly spherical shape
all along the process and no aggregates form. The MLV size
reductions measured when the MLV is in contact with a PE and a
neutral polymer solutions are plotted in the inset of
fig.\ref{FIG:5} and show in both cases a linear variation with
time. The speed is however about $30$ faster for the PE than for
the neutral polymer, although the neutral polymer concentration is
higher and osmotic effects are thus expected to be higher. This
ensures that osmotic effects are negligible in our observations.
Instead, strong electrostatic, presumably coupled to hydrophobic,
attractions between the surfactant bilayers and the PE are the
driving force for the observed transition of the MLV.

\subsubsection{Polyelectrolyte-induced pore formation and growth}

Pore formation in unilamellar vesicles has been observed in
different experimental conditions (application of an electric
field \cite{Pore1,Pore2}, interaction with an amphiphilic protein
\cite{Pore3} or attractive interactions with a surface
\cite{Pore4}). Pores are commonly assumed to be created above a
critical surface tension for the membrane. Electrostatic-driven
tension and rupture of a bilayer has been recently observed when a
charged vesicle is immersed in a solution containing oppositely
charged macroions such as very small vesicles or colloidal
particles \cite{Solon_thesis}. In Ref.\cite{Solon_thesis}, the
tension results from the coating of the macroions by part of the
lipid membrane. In our case, a very similar mechanism comes
presumably into play. In fact, because of the attractive
interactions between the PE and the surfactant bilayer, part of
the surface area of the outside bilayer is used up to form
PE/surfactant complexes. This creates a tension in the bilayer
which ruptures above a critical tension, leading to the formation
of a pore. Generally, the pore growth, stability, and closing
result from a competition between the line tension (of the rim)
and the surface tension of the membrane \cite{Sandre}. Pore growth
is very often arrested as the surface tension relaxes because of
the hole opening and due to the release of inner solvent in the
outside medium. By contrast, in our case, the pore grows until
failure because the surface tension cannot relax faster than the
pore grows. This may be due to the combination of two effects:
first, because of the continuous supply of PE, the surface tension
continues to increase;  secondly, owing to the fact the vesicle is
multilamellar, the release of solvent requires the water to
permeate through the dense stack of bilayers and is very slow, as
inferred from the osmotic de-swelling experiment (inset
fig.\ref{FIG:5}). Examples on unilamellar copolymer vesicles where
pores (formed by application an electric field) grow until failure
can be found in Ref.\cite{Bermudez2003}. We are not aware of
experimental studies visualizing in real-time pore formation and
growth in multilamellar systems. We point out nevertheless that a
peeling mechanism has been invoked for the destabilization of
liposomes in contact with a
 peptide \cite{BBB} involving probably the same kind of
 mechanisms as the one proposed here but Ref.\cite{BBB} does not provide clear
 experimental data nor detail investigation supporting a peeling mechanism.

Very often, when pores open in giant unilamellar vesicles, strong
shape fluctuations due to thermal energy are observed (see for
instance \cite{Solon_thesis,Bermudez2003}). In our experiments,
the MLVs display pronounced shape changes as well, which shows
that shape fluctuations, and pore opening and growth are
intimately connected in multilamellar systems as well. This is
surprising and indicates that cooperative effects involving a very
large number of bilayers come into play. The rapid transitions
between a strongly distorted MLV and a perfectly spherical one
hint at the crucial role of surface tension which leads, in the
absence of pores, the MLV to recover a spherical shape.

The pore growth at constant speed is puzzling. In fact,
exponential growth has been observed experimentally with
unilamellar vesicles, in agreement with theoretical predictions
when friction in the bilayer dominates over viscous dissipation in
the surrounding liquid \cite{Sandre}. By contrast, a linear growth
is predicted in the opposite limit with a velocity $V=\sigma_0 /
\eta_0$
 where $\sigma_0$ is the initial surface tension of the bilayer and $\eta_0$ is the viscosity
of the liquid inside the vesicle. Using the velocity measured
experimentally, and $\eta_0=10^{-3}$ Pas as the viscosity of
water, one obtains that the initial surface tension is of the
order of $10^{-8}$ N/m. This value is extremely small.
Uncertainties may come from the evaluation of the viscosity
$\eta_0$. In fact, a much higher viscosity should be taken into
account due to the packing of bilayers in the MLV which certainly
increases the viscous dissipation.


\section{Conclusions}
\label{SEC:conl}

We have studied mixtures of polyelectrolytes and bilayer-forming
oppositely charged surfactants and have investigated the
morphology and structures of the complexes formed when the two
species are combined together. To better understand the dynamics
for the formation of the surfactant/PE complexes, we have chosen a
novel approach for real-time light microscope visualization of the
interactions of individual mesoscopic surfactant self-assemblies
(multilamellar vesicles) with PE solutions. We have shown that the
PE provokes a peeling of the MLV and that complexes formed
progressively through the peeling-off of the bilayers. Although
real-time light microscopy is not a standard method in the field
of associative precipitation of surfactant and polymers, it is
frequently used with giant and unilamellar lipid vesicles. For
instance, interactions between charged lipids and polyelectrolytes
have been studied and have shown that the polyelectrolyte provokes
endocytosis phenomena \cite{Angelova,Hristova}. Our experiment
with MLVs exhibit a very different behavior. In general, when
multilamellar vesicles are used, richer behaviors may be expected,
since cooperative effects due to the dense packing of bilayers may
play an important role.

Future work in this area should generate a more complete
description of the dynamics of formation of
polyelectrolyte/surfactant complexes and should address the role
of kinetics and osmotic effects. Preliminary work in this
direction shows more complex and intriguing behaviors when the PE
concentration increases.

\bigskip
\bigskip

\textbf{Acknowledgments.} Discussions with M. In and G. Porte are
greatly acknowledged. We thank P. Dieudonn\'e for his assistance
during the X-ray scattering experiments and L. Cipelletti for a
critical reading of the manuscript.


\end{document}